# Full Hydrodynamic Model of Nonlinear Electromagnetic Response in Metallic Metamaterials

Ming Fang, Zhixiang Huang, Wei E. I. Sha, Xiaoyan Y.Z. Xiong, and Xianliang Wu

***Abstract*—Applications of metallic metamaterials have generated significant interest in recent years. Electromagnetic behavior of metamaterials in the optical range is usually characterized by a local-linear response. In this article, we develop a finite-difference time-domain (FDTD) solution of the hydrodynamic model that describes a free electron gas in metals. Extending beyond the local-linear response, the hydrodynamic model enables numerical investigation of nonlocal and nonlinear interactions between electromagnetic waves and metallic metamaterials. By explicitly imposing the current continuity constraint, the proposed model is solved in a self-consistent manner. Charge, energy and angular momentum conservation laws of high-order harmonic generation have been demonstrated for the first time by the Maxwell-hydrodynamic FDTD model. The model yields nonlinear optical responses for complex metallic metamaterials irradiated by a variety of waveforms. Consequently, the multiphysics model opens up unique opportunities for characterizing and designing nonlinear nanodevices.**

## I. Introduction

THE field of metallic metamaterials has demonstrated spectacular experimental progresses in recent years. One of the main driving forces in the field of metallic metamaterials is the potential usage of these materials in nanoantenna, nanosensing, and plasmonic devices [1-5]. In most literature works, the Drude model is commonly used for analyzing the electron gas (plasma) in metals at optical regime; however this approach does not account for the nonlinear and nonlocal electromagnetic responses. Owing to this limitation, the Drude model has been termed the local-response approximation (LRA) model. However, the experimental and fabrication techniques associated with metallic metamaterials are becoming increasingly sophisticated, which allows for an exciting investigations on a very small length scales and/or under the condition of very high-power condition. In these situations, the LRA model with the local response becomes inaccurate and even incorrect. In the regimes of small scale and high power, nonlinear and quantum effects may lead to unusual physical phenomena [6-9]. For example, when an electromagnetic wave strongly interacts with an metallic metamaterial with a versatile geometry, the wave can couple to free electrons near the metal surface resulting in a normal linear response and inherent nonlinear optical behavior [10-12] (generating fundamental and high-order harmonics). These nonlinear optical effects are promising for optimizing metamaterials used in detection, sensing, electromagnetic sources, and quantum information.

A full quantum treatment of the situation, such as density functional theory (DFT), would be the best approach [13]. However, owing to computational limitations and a high number of electrons within the metallic metamaterials, such a fully quantum mechanical model is limited to very small nanoparticles. For "large" plasmonic systems, the quantum approach becomes very demanding. Alternatively, a classical electrodynamic framework with suitable approximations and improvements could be used. For instance, a perturbative treatment of local polarization by the hydrodynamic model allows to obtain the generation of high-order harmonics in metals using a quasi-Fourier transform [14], [15].

Inspired by the above observations, a non-perturbative and universal Maxwell-hydrodynamic model has been proposed to investigate the nonlinear and nonlocal effects. A self-consistent finite-difference time-domain (FDTD) method was used here with the multiphysics model to simulate the interaction between electromagnetic waves and electron gas in metallic nanostructures. The proposed time-domain model fully considers the linear and nonlinear dynamics of the electron gas and does not rely on the experimentally measured bulk and surface nonlinear susceptibilities. The proposed model can also be potentially used for investigating other quantum effects. Compared to the Maxwell-hydrodynamic model in previous literatures, our model has the following differences and advantages. First, our model is based on the classical time-domain FDTD approach and the hydrodynamic equation is solved nonperturbatively. In contrast to the time-domain approach, the alternative frequency-domain approaches such as boundary element method (BEM) and finite element method

Manuscript received XX XX, 2015; revised XX XX, 2015; accepted XX XX, 2015. This work was supported the NSFC (Nos. 61101064, 51277001 and 61201122), Universities Natural Science Foundation of Anhui Province (No. KJ2012A013), DFMEC (No. 20123401110009) and NCET (NCET-12-0596) of China.

M. Fang, Z. Huang and X. Wu are with the Key Laboratory of Intelligent Computing and Signal Processing, Ministry of Education, Anhui University, Hefei 230039, China. E-mail: zxhuang@ahu.edu.cn (Z.X. Huang)

M. Fang, W. E. I. Sha, and X. Y. Z. Xiong are with the Department of Electrical and Electronic Engineering, The University of Hong Kong. E-mail: wsha@eee.hku.hk (W.E.I. Sha); xiong913@gmail.com (X.Y.Z. Xiong) (FEM) [16,17] compute the nonlinear response relying on the results of perturbation theory; and the density of electrons is treated as homogeneous. Second, specified interpolation techniques are proposed, which is essential to capture the fundamental conservation laws of charge, energy and angular momentum governing the high-order harmonic generation. Third, different from other models [18, 19] showing some inconsistences, our model produces consistent results with analytical models in literatures [20]. Four, both linearly and circularly polarized excitations are successfully implemented, which is important to understand versatile selection rules of second-harmonic generation (SHG). Finally, the numerical aspects of the proposed model including accuracy, convergence and stability are analyzed in detail.

This paper is structured as follows. The proposed approach's numerical framework is introduced in Section II. In Section III, the linear response of metamaterials is first calculated and the result is compared with the classical Drude model. Then, nonlinear responses of a standard gold nanosphere and a variety of metasurfaces are presented. The numerical results demonstrating that the Maxwell-hydrodynamic model can be used for simulating the high-order harmonics in metallic metamaterials. Moreover, the model also demonstrates a powerful potential for investigating the characteristics of second harmonic radiation for the metasurfaces with different symmetries.

## II. Computational Method

### A. Electromagnetic-Hydrodynamic Model

The interaction of electromagnetic fields **E** and **H** with an arbitrary nonmagnetic (metallic) material can be described by

$$\nabla \times \mathbf{H} = \varepsilon_0 \frac{\partial \mathbf{E}}{\partial t} + \frac{\partial \mathbf{P}}{\partial t} \quad (1)$$

$$\nabla \times \mathbf{E} = -\mu_0 \frac{\partial \mathbf{H}}{\partial t}, \quad (2)$$

where $\varepsilon_0$ and $\mu_0$ are the vacuum permittivity and permeability, **P** is the polarization. To investigate the dynamics of electrons in a metallic/plasmonic system, the hydrodynamic model [21] is adopted

$$\frac{\partial \mathbf{v}}{\partial t} + \mathbf{v}\cdot\nabla\mathbf{v} = -\frac{e}{m}(\mathbf{E} + \mu_0 \mathbf{v}\times\mathbf{H}) - \gamma\mathbf{v} - \frac{\nabla p}{n} \quad (3)$$

$$\frac{\partial n}{\partial t} = -\nabla\cdot(n\mathbf{v}) \quad , \quad (4)$$

where $n(\mathbf{r},t)$ and $\mathbf{v}(\mathbf{r},t)$ are the time- and position-dependent electron density and velocity, $\gamma$ is the phenomenological damping frequency (capturing optical loss), $e$ and $m$ are the electron charge and mass, respectively, and $p$ is the quantum pressure $p = (3\pi^2)^{2/3}(\hbar/5)n^{5/3}$, evaluated according to the Thomas-Fermi theory [22]. Equation (3) is an Euler's equation representing the force balance on a fluid element, while equation (4) is the current continuity equation, ensuring the conservation of charge. The nonlinear terms $\mathbf{v}\cdot\nabla\mathbf{v}$, $\mathbf{v}\times\mathbf{H}$, and $n\mathbf{v}$ underlie the physical origins of the nonlinear and nonlocal effects. The quantum pressure term can be ignored for metallic metamaterials with a typical size larger than 10 nm. Maxwell's equations (1-2) and hydrodynamic equations (3-4) are coupled through the polarization term **P**, which can be expressed as the velocity

$$\frac{\partial \mathbf{P}}{\partial t} = -en\mathbf{v}. \quad (5)$$

Equations (1)-(5) provide a self-consistent formulation of free electron gas in plasmonic systems. Both nonlocal and nonlinear effects can be considered using the self-consistent formulations. The multiphysics equations can be numerically solved by employing the FDTD method [23]-[27], which has been extended to model nonlinear and nonlocal effects in dispersive media.

### B. The Computational Grids

To numerically model the transient problem in the plasmonic systems, we used the Yee grids to maintain charge conservation and divergence-free condition as shown in Fig. 1. The electric field **E** and electron density n are defined at the time step $l$+1/2 and are respectively located at the grid face center and grid center, respectively. The velocity **v** and magnetic field **H** are defined at the time step $l$ and located at the grid face center and the grid edges, respectively.

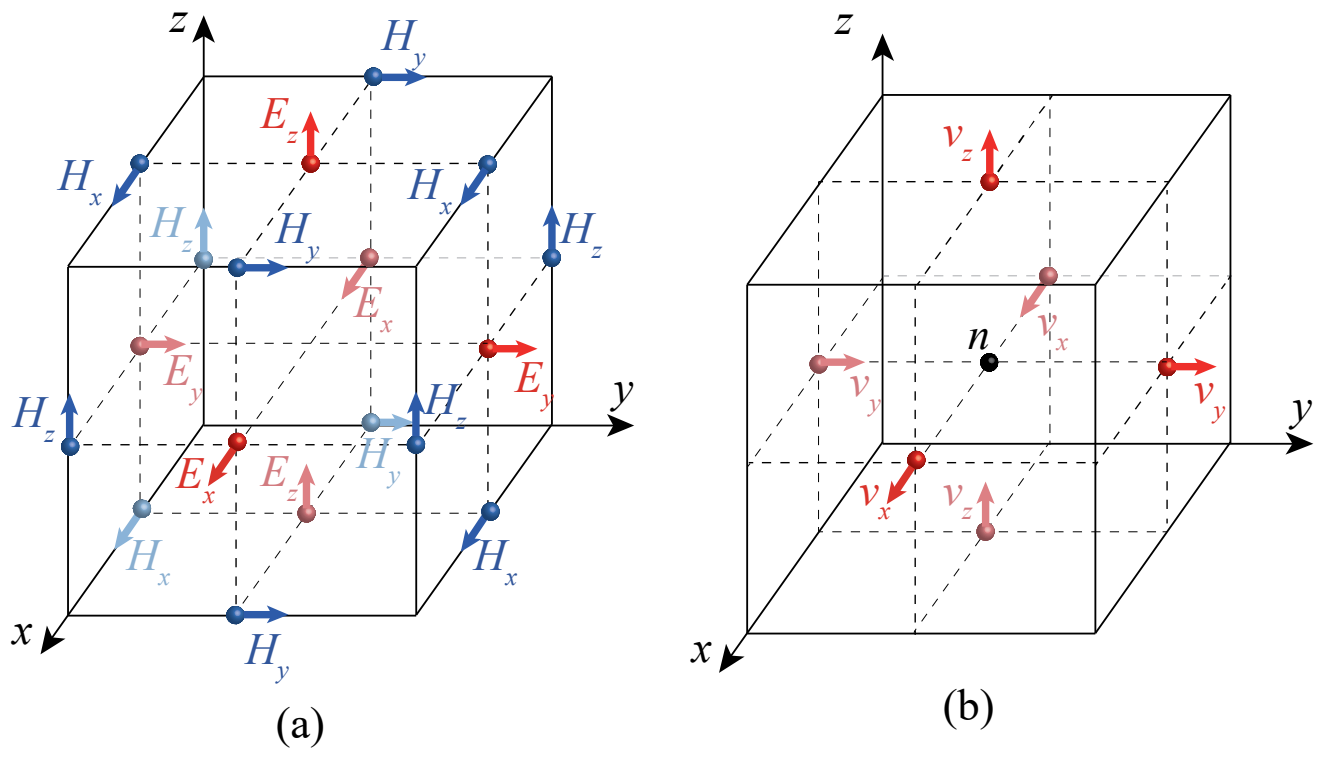

Fig. 1. The Yee grids for the Maxwell-hydrodynamic system. (a) Maxwell's equations. (b) Hydrodynamic equations.

With the central spatial difference and with the explicit time stepping, the equations for updating the $x$ components of the physical quantities of **E**, **H**, **v**, and $n$ are given by

$$E_x^{l+1/2}(i,j+1/2,k+1/2)=E_x^{l-1/2}(i,j+1/2,k+1/2)+\frac{\Delta t}{\varepsilon_0\Delta y}\left(H_z^l(i,j+1,k+1/2)-H_z^l(i,j,k+1/2)\right)$$
$$-\frac{\Delta t}{\varepsilon_0\Delta z}\left(H_y^l(i,j+1/2,k+1)-H_y^l(i,j+1/2,k)\right)+\frac{\Delta te}{\varepsilon_0}\bar{n}^l(i,j+1/2,k+1/2)\cdot v_x^l(i,j+1/2,k+1/2) \quad (6)$$

$$H_x^{l+1}(i+1/2,j,k)=H_x^l(i+1/2,j,k)+\frac{\Delta t}{\mu_0\Delta z}\left(E_y^{l+1/2}(i+1/2,j,k+1/2)-E_y^{l+1/2}(i+1/2,j,k-1/2)\right) \quad (7)$$
$$-\frac{\Delta t}{\mu_0\Delta y}\left(E_z^{l+1/2}(i+1/2,j+1/2,k)-E_z^{l+1/2}(i+1/2,j-1/2,k)\right)$$

$$v_x^{l+1}(i,j+1/2,k+1/2)=v_x^l(i,j+1/2,k+1/2)-\Delta t\gamma v_x^l(i,j+1/2,k+1/2)-\frac{\Delta te}{m}E_x^{l+1/2}(i,j+1/2,k+1/2)$$
$$-\left[\frac{\Delta t}{2\Delta x}\left(v_x^l(i,j+1/2,k+1/2)\cdot v_x^l(i+1,j+1/2,k+1/2)-v_x^l(i,j+1/2,k+1/2)\cdot v_x^l(i-1,j+1/2,k+1/2)\right)\right.$$
$$+\frac{\Delta t}{2\Delta y}\left(\bar{v}_y^l(i,j+1/2,k+1/2)\cdot v_x^l(i,j+3/2,k+1/2)-\bar{v}_y^l(i,j+1/2,k+1/2)\cdot v_x^l(i,j-1/2,k+1/2)\right) \quad (8)$$
$$\left.+\frac{\Delta t}{2\Delta z}\left(\bar{v}_z^l(i,j+1/2,k+1/2)\cdot v_x^l(i,j+1/2,k+3/2)-\bar{v}_z^l(i,j+1/2,k+1/2)\cdot v_x^l(i,j+1/2,k-1/2)\right)\right]$$
$$-\frac{\Delta t\mu_0 e}{m}\left(\bar{v}_y^l(i,j+1/2,k+1/2)\bar{H}_z^{l+0.5}(i,j+1/2,k+1/2)-\bar{v}_z^l(i,j+1/2,k+1/2)\cdot\bar{H}_y^{l+0.5}(i,j+1/2,k+1/2)\right)$$

$$n^{l+3/2}(i+1/2,j+1/2,k+1/2)=n^{l+1/2}(i+1/2,j+1/2,k+1/2)$$
$$-\left[\frac{\Delta t}{\Delta x}\left(\bar{n}^{l+1/2}(i+1,j+1/2,k+1/2)v_x^{l+1}(i+1,j+1/2,k+1/2)-\bar{n}^{l+1/2}(i,j+1/2,k+1/2)v_x^{l+1}(i,j+1/2,k+1/2)\right)\right. . \quad (9)$$
$$+\frac{\Delta t}{\Delta y}\left(\bar{n}^{l+1/2}(i+1/2,j+1,k+1/2)v_y^{l+1}(i+1/2,j+1,k+1/2)-\bar{n}^{l+1/2}(i+1/2,j,k+1/2)v_y^{l+1}(i+1/2,j,k+1/2)\right)$$
$$\left.+\frac{\Delta t}{\Delta z}\left(\bar{n}^{l+1/2}(i+1/2,j+1/2,k+1)v_z^{l+1}(i+1/2,j+1/2,k+1)-\bar{n}^{l+1/2}(i+1/2,j+1/2,k)v_z^{l+1}(i+1/2,j+1/2,k)\right)\right]$$

where the overbar notations correspond to spatial and temporal averaging. For example, the quantity $\bar{H}_y^{l+1/2}(i,j+1/2,k+1/2)$ is the spatial and time average of $H_y^{l+1}(i,j+1/2,k)$, $H_y^{l+1}(i,j+1/2,k+1)$, $H_y^l(i,j+1/2,k)$, and $H_y^l(i,j+1/2,k+1)$. The discrete equations for the $y$ and $z$ components can be derived in a similar way.

### C. *Stability Condition of the Method*

In general, the stability of the FDTD scheme can be analyzed by finding the roots of the corresponding growth matrix. Regarding the FDTD method, the stability condition obeys the Courant-Friedrichs limit (CFL)

$$\Delta t \le \frac{1}{c}\left(\sqrt{\frac{1}{(\Delta x)^2}+\frac{1}{(\Delta y)^2}+\frac{1}{(\Delta z)^2}}\right)^{-1}. \quad (10)$$

Unfortunately, the matrix roots for the coupled Maxwell-hydrodynamic system are difficult to obtain in a closed form [28]. The system amplification matrices also cannot be trivially found or defined for the nonlinear coupled system including the four variables $\{\mathbf{E},\mathbf{H},\mathbf{v},n\}$ and their products, such as $n\mathbf{v}$, $\mathbf{v}\cdot\nabla\mathbf{v}$ and $\mathbf{v}\times\mathbf{H}$. Here, the nonlinear terms of $\mathbf{v}\cdot\nabla\mathbf{v}$ and $\mathbf{v}\times\mathbf{H}$ in Eq. (3) mainly contribute to the high-order harmonic generation. Compared to the linear response, the nonlinear response from centrosymmetric metals is much weaker. Therefore, the amplitudes of the linear (fundamental) scattered waves in the nonlinear model is regarded to be the same as that of the scattered waves in a linear Drude model, which is called the undepleted pump approximation. After removing the two nonlinear terms $\mathbf{v}\cdot\nabla\mathbf{v}$ and $\mathbf{v}\times\mathbf{H}$ in Eq. (3), we have

$$\frac{\partial\mathbf{v}}{\partial t}+\gamma\mathbf{v}=-\frac{e}{m}\mathbf{E} \quad (11)$$

The above is nothing but the linear Drude model for metals. After modeling many complex metallic nanostructures under a long-term simulation, we found that the hydrodynamic model was stable within the commonly used CFL condition in the FDTD method. Physically, the nonlinear terms, which produce much weaker nonlinear response than the linear one, do not significantly affect the stability condition of the FDTD method with the linear Drude model.

## III. Numerical Results

### A. Linear Response

In a recent paper, we outlined a computational method for investigating the loss compensation and photoluminescence spectra of the metamaterials and a four-level atomic gain media coupled system [29]. The Drude model was used to simulate the plasmonic metamaterials, and can only be used for investigating the linear and local properties of these materials. In this section, the developed self-consistent Maxwell-hydrodynamic equations are utilized for calculating the optical properties of plasmonic metamaterials. The parameters used were the same as the parameters in [29]. The initial electron density $n_0$ was obtained from the plasma frequency $\omega_p$ of the Drude model $n_0 = \varepsilon_0 m \omega_p^2 / e^2$. We computed the transmission spectra of a metasurface, and validated the self-consistent model by comparing it to the Drude model, and the results are shown in Fig. 2 (a). Here, we show the results for a periodic array of complementary split ring resonators (CSRRs) with a lattice constant of $D = 540$ nm sandwiched between a glass substrate and a polymethyl methacrylate (PMMA) layer. Figure 2(b) shows the feature size of the unit cell of CSRR. The thicknesses of the PMMA layer and the glass slab were 180 nm and 50 nm, with the relative permittivity of 2.2 and 2.56, respectively.

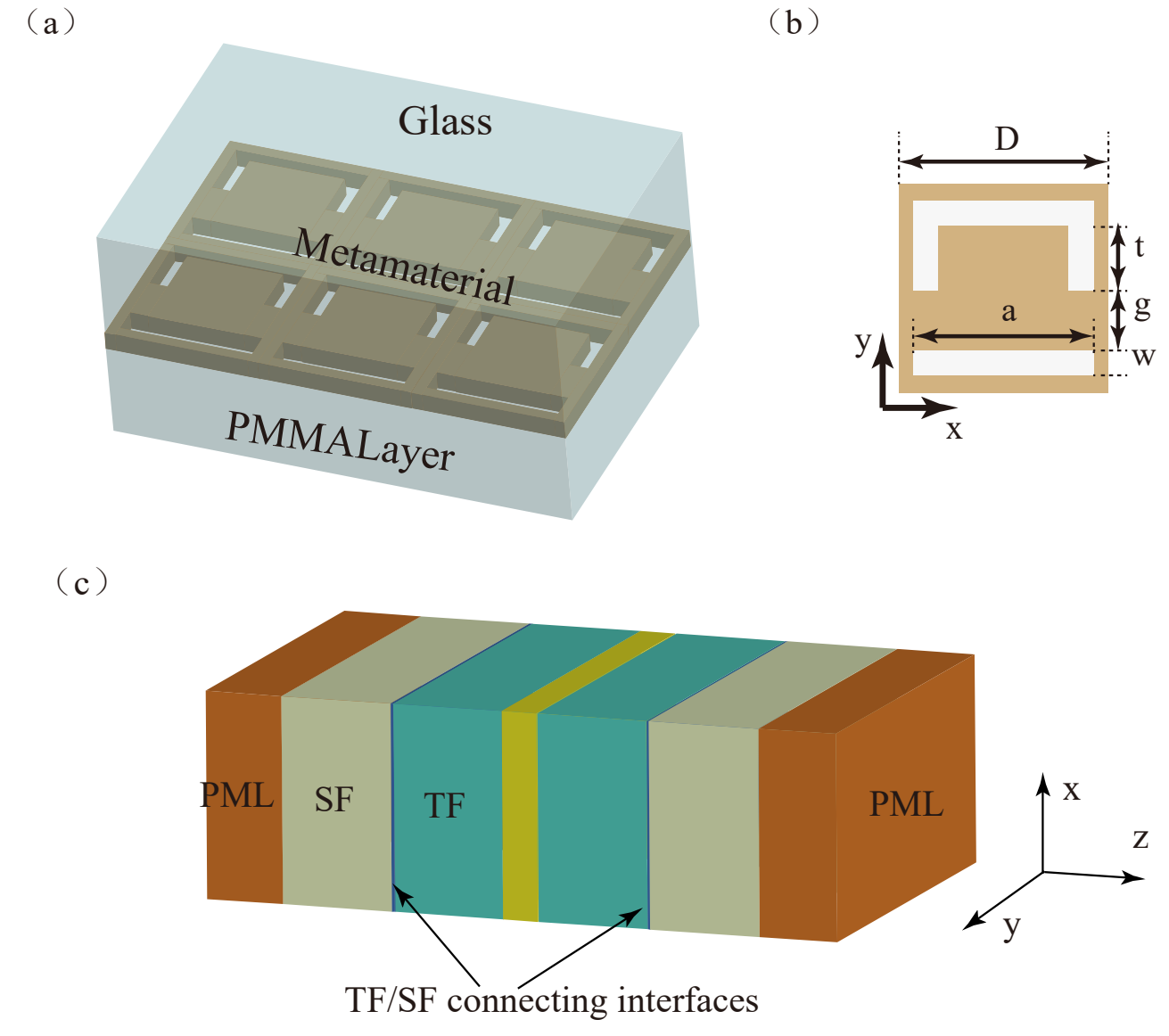

Fig. 2. (a) Schematic pattern of the metamaterial comprising a periodic array of complementary SRRs. (b) The feature size of a unit cell. D = 540 nm, lateral slit a = 470 nm, top vertical slit and gap t = g = 170 nm, and slit width w = 65 nm. (c) Configuration for the FDTD method. PMLs were set to be in the *z* direction and periodic boundary conditions were used at the both *x* and *y* directions.

In the FDTD scheme, the probe source was an *x*-polarized or *y*-polarized Gaussian pulse signal, propagating in the *z*-direction. The source was introduced using the total-field and scattered-field (TF/SF) technique [30]. The configuration for the FDTD method is illustrated in Fig. 2(c). Perfectly matched layers (PMLs) were set to be in the *z* direction and periodic boundary conditions are employed in both the *x* and *y* directions. The temporal evolutions of the transmitted and reflected fields were recorded in the two planes in the total field region and scattered field region, respectively. Next, the data were Fourier-transformed for determining the complex transmission $t(\omega)$ and reflection $r(\omega)$ coefficients in the frequency domain. The results were calculated by adopting uniform spatial steps $\Delta x = \Delta y = \Delta z$. The convergence of the model was validated by using different spatial steps, from $5\times10^{-9}$ m through $2.5\times10^{-9}$ m to $1\times10^{-9}$ m.

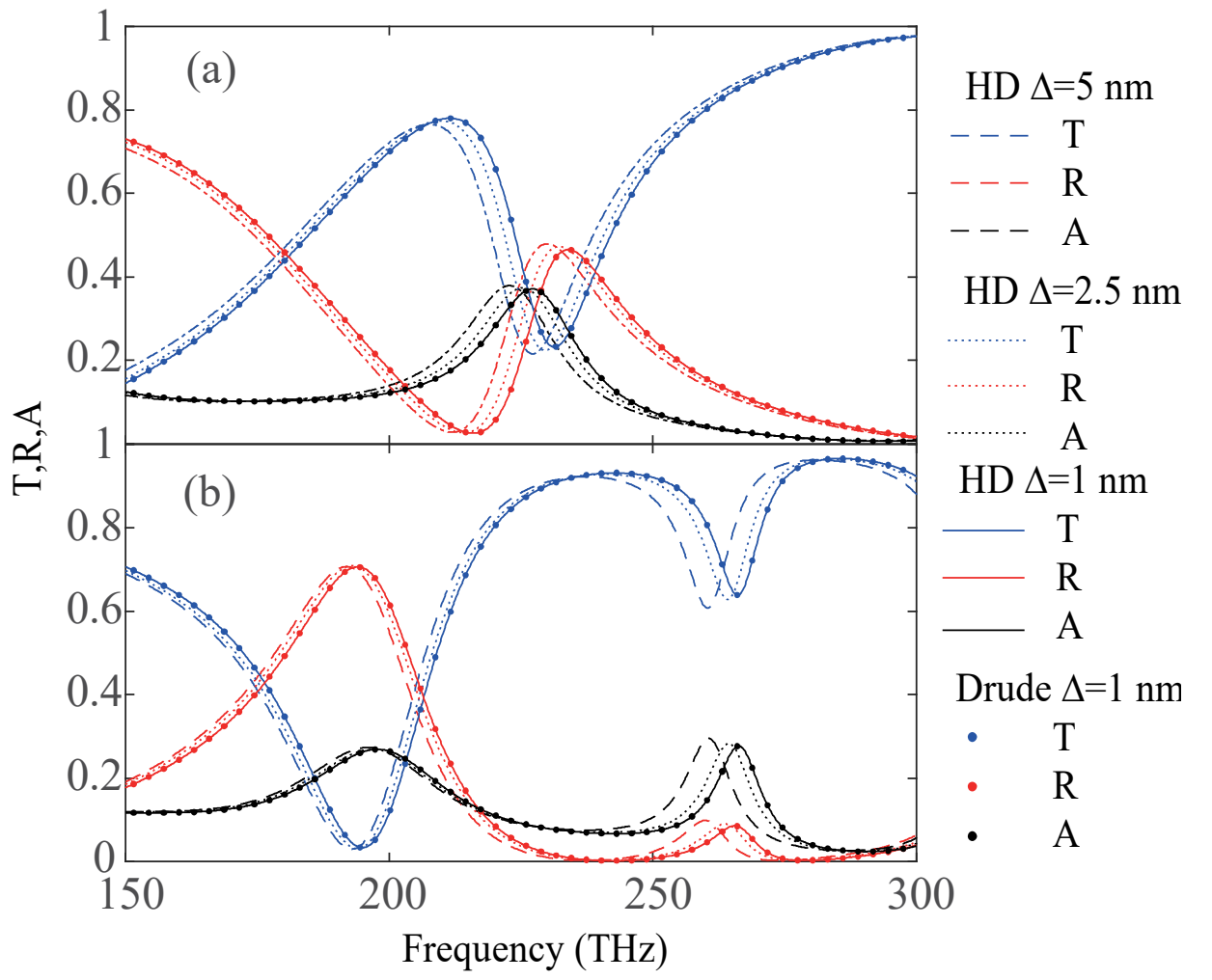

Fig. 3 The calculated spectra of transmittance ($T(\omega)=|t(\omega)|^2$), reflectance ($R(\omega)=|r(\omega)|^2$), and absorptance ($A(\omega)=1-T(\omega)-R(\omega)$) for the metamaterial system in Fig. 2. The results are shown for (a) *x*-polarization and (b) *y*-polarization.

In Fig. 3, we show the simulated transmittance $T(\omega)=|t(\omega)|^2$, reflectance $R(\omega)=|r(\omega)|^2$, and absorptance $A(\omega)=1-T(\omega)-R(\omega)$ spectra as functions of frequency. In Fig. 3, dotted lines indicate the spectra obtained using the Drude model, while the other lines are the spectra obtained using the Maxwell-hydrodynamic model with different spatial steps. One can see that the results obtained using the proposed method are in good agreements with those obtained using the Drude method. In addition, the results obtained using our method are comparable to the experimental results in [31].

### B. Characteristics of Nonlinear Response

In this section, we study the nonlinear response from a metallic nanoparticle to demonstrate that our model is a rigorous and universal approach. As a case study, we consider the nonlinear scattering from a 20-nm-radius ($R = 20$ nm) gold sphere, illuminated by a cosine-modulated Gaussian pulse. The parameters of gold were chosen as $n_0 = 5.98\times10^{28}$ m$^{-3}$ and $\gamma = 1.075\times10^{14}$ s$^{-1}$. The three-dimensional computational domain is shown in Fig. 4; in this system, the PMLs were imposed in the $x$, $y$, and $z$ directions. The incident pulse was an $x$-polarized plane wave that propagates in the $z$ direction, with a fundamental carrier frequency $f_0 = 577$ THz and a temporal width $\tau = 10$ fs. The settings of the pulse were chosen for ensuring significant separation of second and third harmonic spectra from fundamental spectra. The maximal peak intensity of $I_0 = 9\times10^{18}$ W/m$^2$ was selected to ensure a stable and significant nonlinear process. The spatial and temporal steps were $\Delta x = \Delta y = \Delta z = 1.25\times10^{-9}$ m and $\Delta t = 1.5\times10^{-18}$ s. The temporal offset was $t_0 = 3\tau$ and total simulation time was $T = 7\tau$.

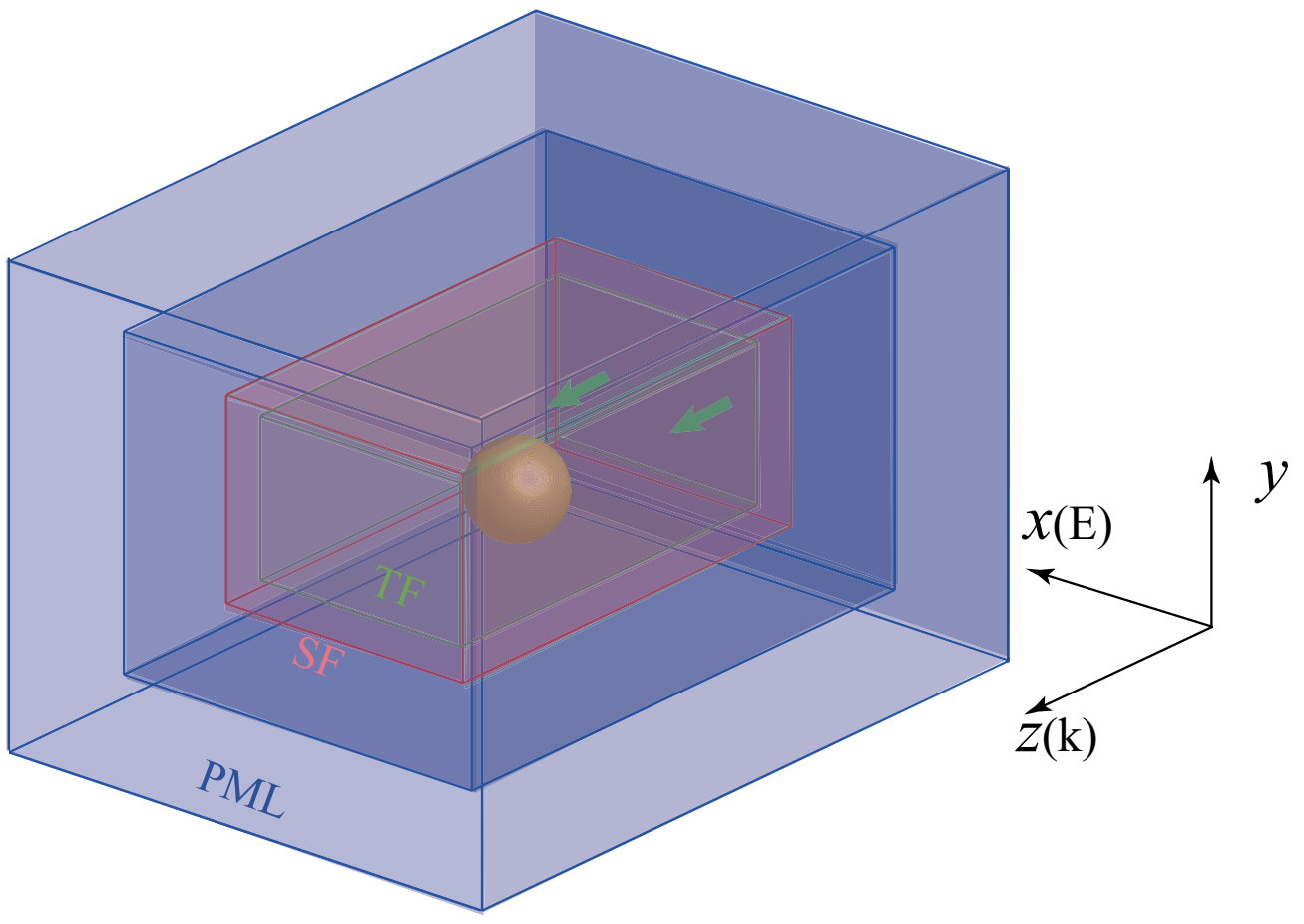

Fig. 4. Layout of the simulation of nonlinear optical scattering from a gold nanosphere illuminated by a strong lasing pulse.

Figure 5 shows the $E_x$ component of the scattered field and the fluctuating electron intensity $\Delta N = n - n_0$ on the nanosphere surface, as a function of time. We measured the $E_x$ component of the scattered field at a blue probe point, as shown in the bottom inset of Fig. 5. Meanwhile, the electron density on the surface of the gold nanosphere was recorded at the red probe point, as shown in the bottom inset of Fig. 5. In the nonperturbative model, the total electron number is conserved with time, as shown in the top inset of Fig. 5. The total electron $N_{total}$ is the sum over the electron density on all of the Yee grids within the nanosphere region.

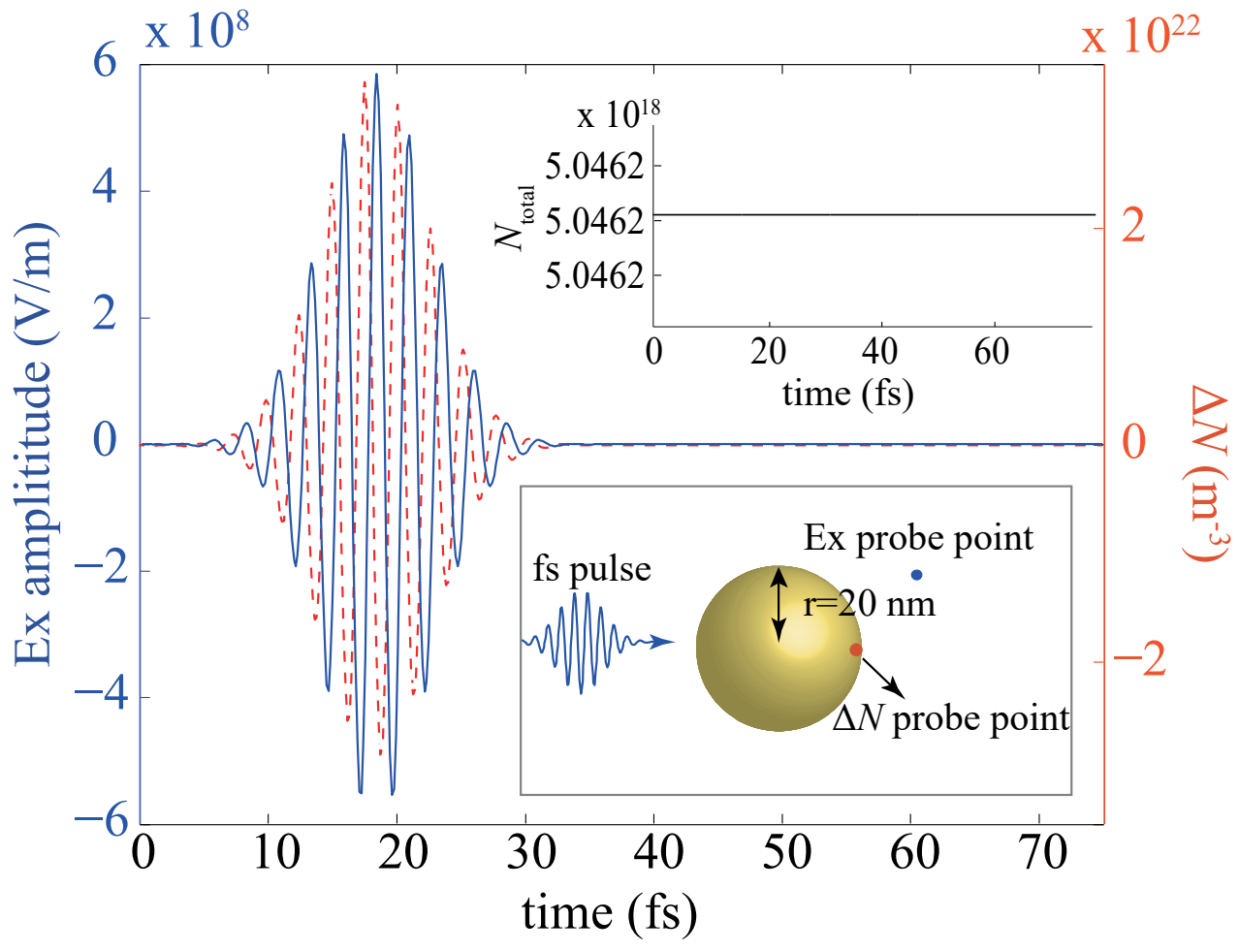

Fig. 5. The $E_x$ component of scattered field (blue solid line) and the fluctuation of charge density on the nanosphere surface (dashed red line), plotted vs. time. The upper inset shows the temporal evolution of the overall electron charges within the nanosphere. The bottom inset schematically shows the simulation configuration for modeling the nonlinear generation of harmonics for the gold nanosphere.

To characterize the generation of high-order harmonics, we Fourier-transformed the time-dependent scattered field in Fig. 5. Both second-harmonic and third-harmonic spectra were detected, as shown in Fig. 6. To further validate our method, we varied the fundamental pump power to investigate the power-dependent nonlinear harmonics. The peak intensities of the second and third harmonics are plotted as functions of the quadratic and cubic power of the fundamental intensity, respectively (Fig. 7). Figure 7 demonstrates that the intensity of the nonlinear harmonics can be satisfactorily described by the power dependence on the fundamental intensity, with second and third power dependence for the second and third harmonics, respectively. Our model captures a correct energy conversion relation between the fundamental pump and high-order harmonics.

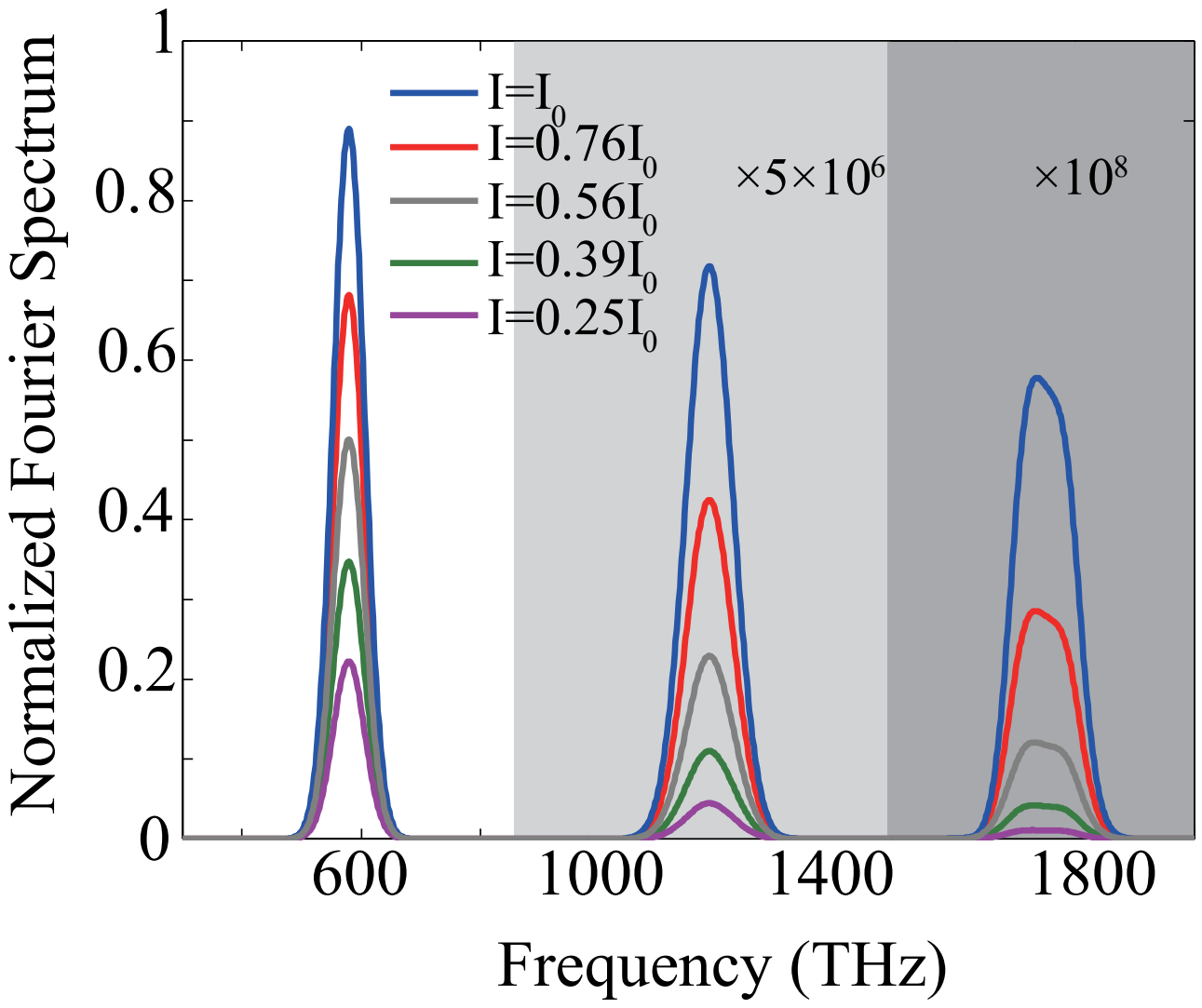

Fig. 6. Nonlinear scattering spectra for a 20 nm radius gold nanosphere, under different pump power.

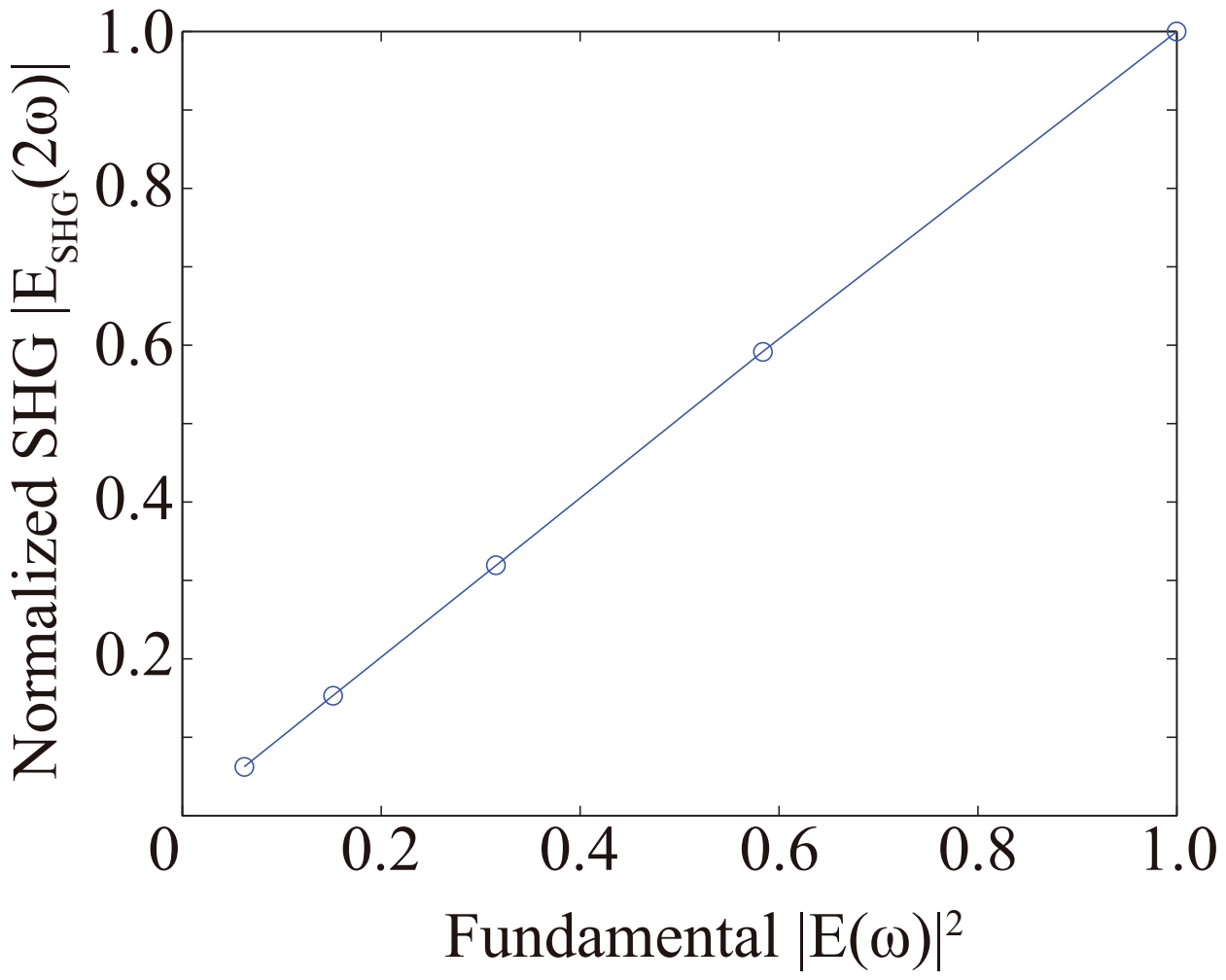

(a)

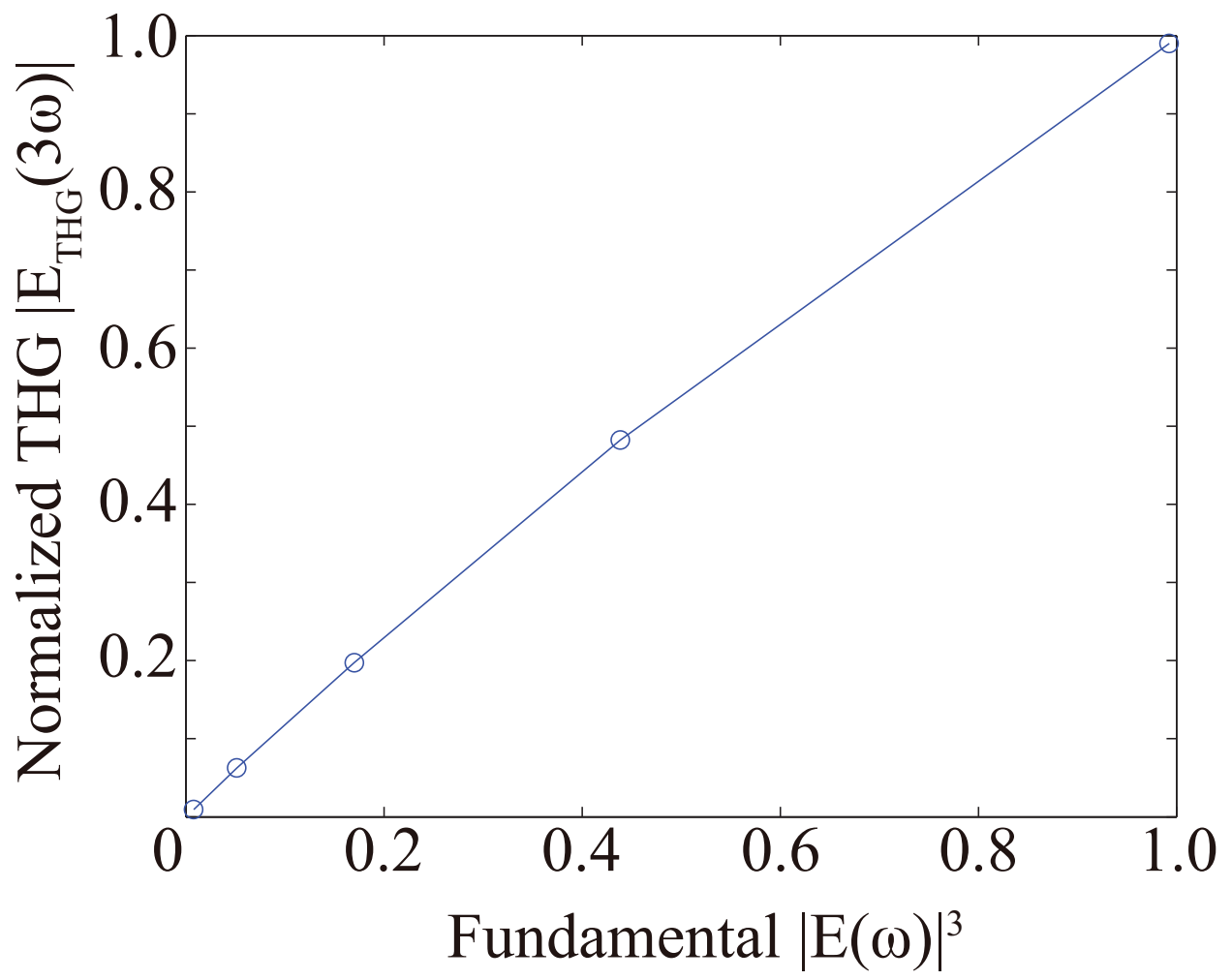

(b)

Fig. 7. The peak intensities of the second and third harmonics vs. the second and third power of the fundamental intensities, respectively. (a) Second harmonics; (b) Third harmonics.

To determine the accuracy and convergence of our method for modeling the second harmonic waves, the radiation pattern corresponding to the second harmonic radiation of the gold sphere in Fig. 5 was also calculated using the near-to-far-field transformation technique [30] and compared to the corresponding analytical solution [20]. Spatial increments were uniform and were 10 nm, 5 nm, 2.5 nm, and 1.25 nm. Figure 8 shows the calculated diagram for the second harmonic radiation, for different scenarios of spatial step, as well as the analytical solution (black curve), at a frequency $f$ = 770 THz along the E-plane. It is worth noting that the linear radiation power in the forward and backward directions is the strongest, as seen in the right inset of Fig. 8. However, the second-harmonic radiation power vanishes in both the forward and backward directions. The unusual angular response of second harmonics is a direct consequence of the rotational symmetry of the sphere with respect to the propagation axial of the fundamental incident wave.

Besides the hydrodynamic model, boundary element method (BEM) is an alternative approach to model the SHG from metals, which has been developed by our group [17]. In contrast to the hydrodynamic model that only needs two input parameters (initial electron density and damping frequency), the BEM needs the surface nonlinear susceptibility of metals relying on the experimental data. At **Appendix A**, we show that the hydrodynamic model results agree with the BEM well. The main deviations between the two methods lie at the side lobes of the second-harmonic radiation patterns. Because for the centrosymmetric metals, the BEM only considers the dominated surface nonlinear sources and ignores the small bulk nonlinear sources for the SHG.

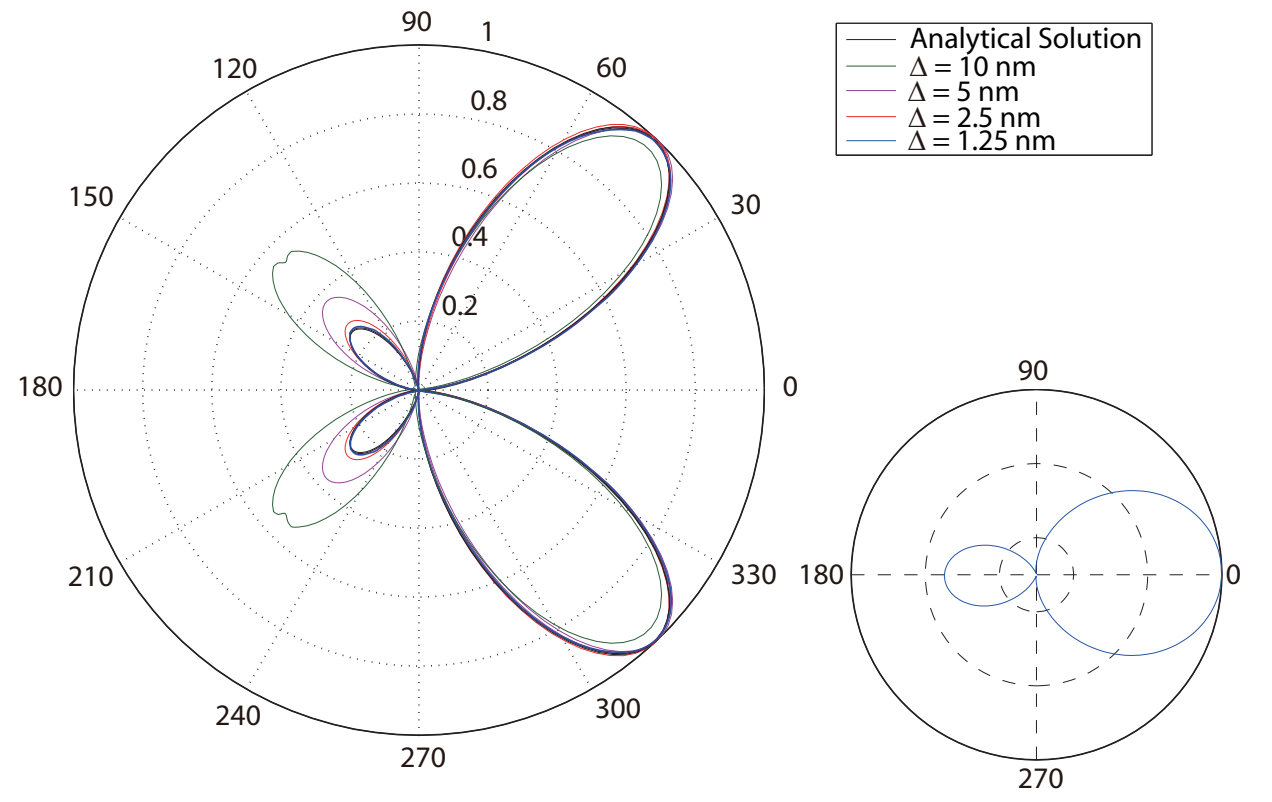

Fig. 8. Second harmonic radiation diagrams as a function of $\theta$ at $\varphi = 0^o$, for a gold nanosphere of radius $R$ =100 nm. The pump wavelength was $\lambda$ = 780 nm and all curves were normalized by the maximum obtained for the result with Δ = 1.25 nm. The right inset shows the radiation pattern at the fundamental frequency.

We used two error criteria to evaluate the accuracy; these criteria were the mean absolute error and the global relative $L_2$ error. The errors, the required memory and the CPU time are listed in Table 1. As shown in the Table 1, higher numerical precision was obtained for smaller spatial steps. The order of convergence $p$ was estimated as

$$p = \ln\left[\frac{Err(h_i) - Err(h_{i+1})}{Err(h_{i+1}) - Err(h_{i+2})}\right] \Big/ \ln\left[\frac{h_i}{h_{i+1}}\right] \tag{12}$$

where $h_i$ is the spatial step. At least three computations were required where $h_i/h_{i+1} = h_{i+1}/h_{i+2}$. The orders of convergence for the mean absolute error were $p$ = 1.3123 (Δ =10 nm, 5 nm, and 2.5 nm) and 1.2444 (Δ =5 nm, 2.5 nm, and 1.25 nm). For the global relative $L_2$ error are 1.077 (Δ =10 nm, 5 nm, and 2.5 nm) and 1.2006 (Δ =5 nm, 2.5 nm, and 1.25 nm). The orders of convergence for both mean absolute error and the global relative $L_2$ error were above 1.

TABLE I Comparisons of computational statistics across different spatial of scenarios discretization

| Δ | Memory (MB) | Time (s) | Time step | Grids | $Err_{L2}$ | $Err_{mean}$ |
|---|---|---|---|---|---|---|
| 10 nm | 756 | 191 | $1\times10^4$ | $80^3$ | 0.2037 | 0.0612 |
| 5 nm | 952 | 682 | $2\times10^4$ | $100^3$ | 0.0912 | 0.0259 |
| 2.5 nm | 1536 | 2364 | $4\times10^4$ | $160^3$ | 0.0379 | 0.0110 |
| 1.25 nm | 3128 | 12884 | $8\times10^4$ | $220^3$ | 0.0151 | 0.0050 |

.

## *C. Perfectly Matched Layer for Higher-Order Harmonics*

The PML technique is an efficient tool to solve "open" problems and absorb outgoing waves at a broadband. Here, we used the split-field PML technique that was proposed by J. P. Berenger [32]. In this technique, each component of electromagnetic fields is split into two subcomponents in the absorbing boundary regions. In our 3D simulation, the six electric and magnetic field components yield 12 subcomponents in the Cartesian coordinates. The discretized $E_{xz}$ subcomponent in the PML region was

$$E_{xz}^{l+1/2}(i,j+1/2,k+1/2)=e^{(-\sigma_y\Delta t/\varepsilon_0)}E_{xz}^{l-1/2}(i,j+1/2,k+1/2)+\frac{1-e^{(-\sigma_y\Delta t/\varepsilon_0)}}{\sigma_y\Delta y}\left(H_z^l(i,j+1,k+1/2)-H_z^l(i,j,k+1/2)\right). \quad (13)$$

Here, $\sigma_y$ is the local electric conductivity at ($i$, $j$+1/2, $k$+1/2) in the PML region. To demonstrate the absorbing property for a nonlinear beam, we considered electromagnetic wave scattered by a 400-nm-radius gold sphere (the material parameters of gold were set as before). The sphere was illuminated by a point source. A Gaussian pulse at the driving frequency of 385 THz was generated by the point source, with a temporal width of 6 fs and electric field peak amplitude of $E_{0,max}$ = $1.0\times10^{13}$ V/m. The parameters of the point source were chosen to obtain satisfactory separation between the scattered linear and nonlinear harmonics in the frequency domain. The space was discretized by employing a FDTD lattice with $\Delta x$ = $\Delta y$ = $\Delta z$ =20 nm, the time step was $\Delta t$ = $3\times10^{-17}$ s and total time steps $N$ = 2000. Ten-cell-thick PML layers were at the grid edges and were placed at a lattice distance of ten cells from the sphere, on all sides. This results in a 80×80×80 cell lattice. The electric field was recorded at a predefined point as shown in Fig. 9.

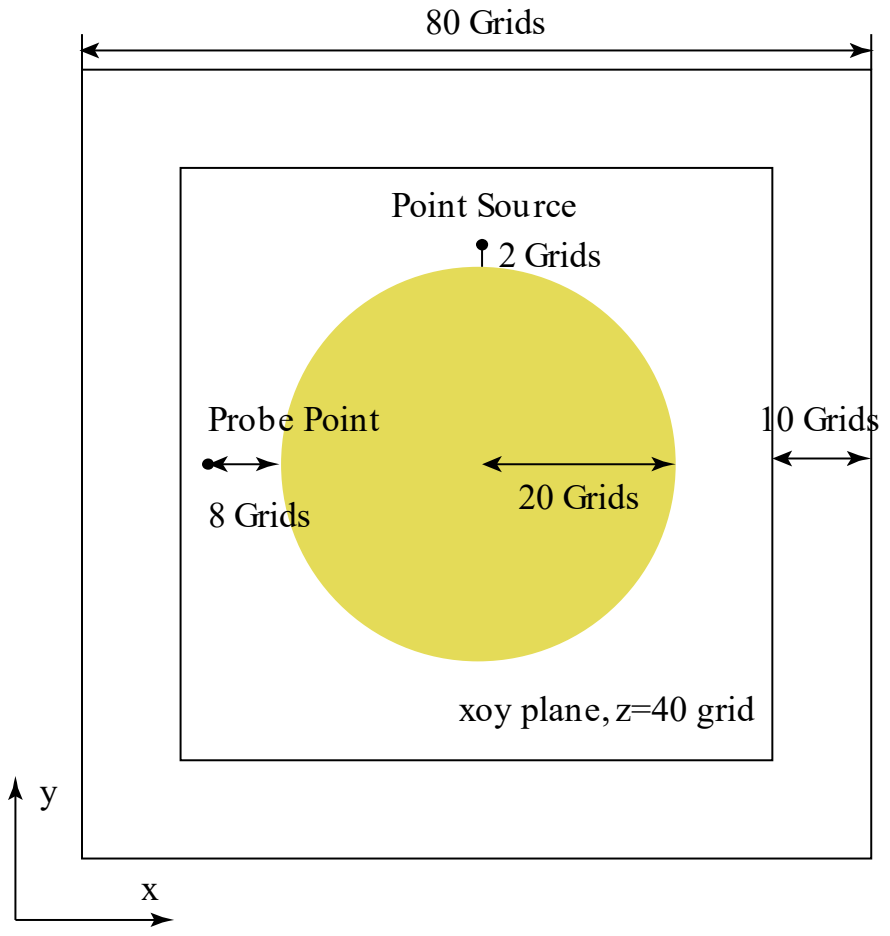

Fig. 9. The FDTD simulation setup. The electric field was recorded at the probe point.

To investigate the nonlinear reflection error owing to the PML, we performed a reference simulation. In this simulation, the mesh was extended to 1920 cells at both the $x$, $y$ and z directions, resulting in a 2000×2000×2000 cell lattice in a system without the PML. The fields were then excited using the identical source. The time-dependent fields are recorded at the same point before waves were reflected back by the truncated boundaries. The error relative to the reference solution was

$$\eta = 20\log_{10}\frac{|\tilde{\chi}-\tilde{\chi}_{ref}|}{\max|\tilde{\chi}_{ref}|} \quad (14)$$

where $\chi$ is the frequency-domain field including the reflection error by the PMLs, $\chi_{ref}$ is the corresponding reference field excluding the reflection error, and max denotes the maximal value.

The electric field profiles in the time domain and in the frequency domain are presented in Figs. 10 (a) and (b). The reflection errors by Eq. (14) are illustrated in Figs. 11 (a-c), respectively for the fundamental harmonics, second harmonics, and third harmonics around their central frequencies. It is demonstrated that the maximum error around the fundamental frequency is on the order of –63 dB. Around the central frequencies of the second and third harmonics, the errors are on the order of -41 dB and -27 dB, respectively.

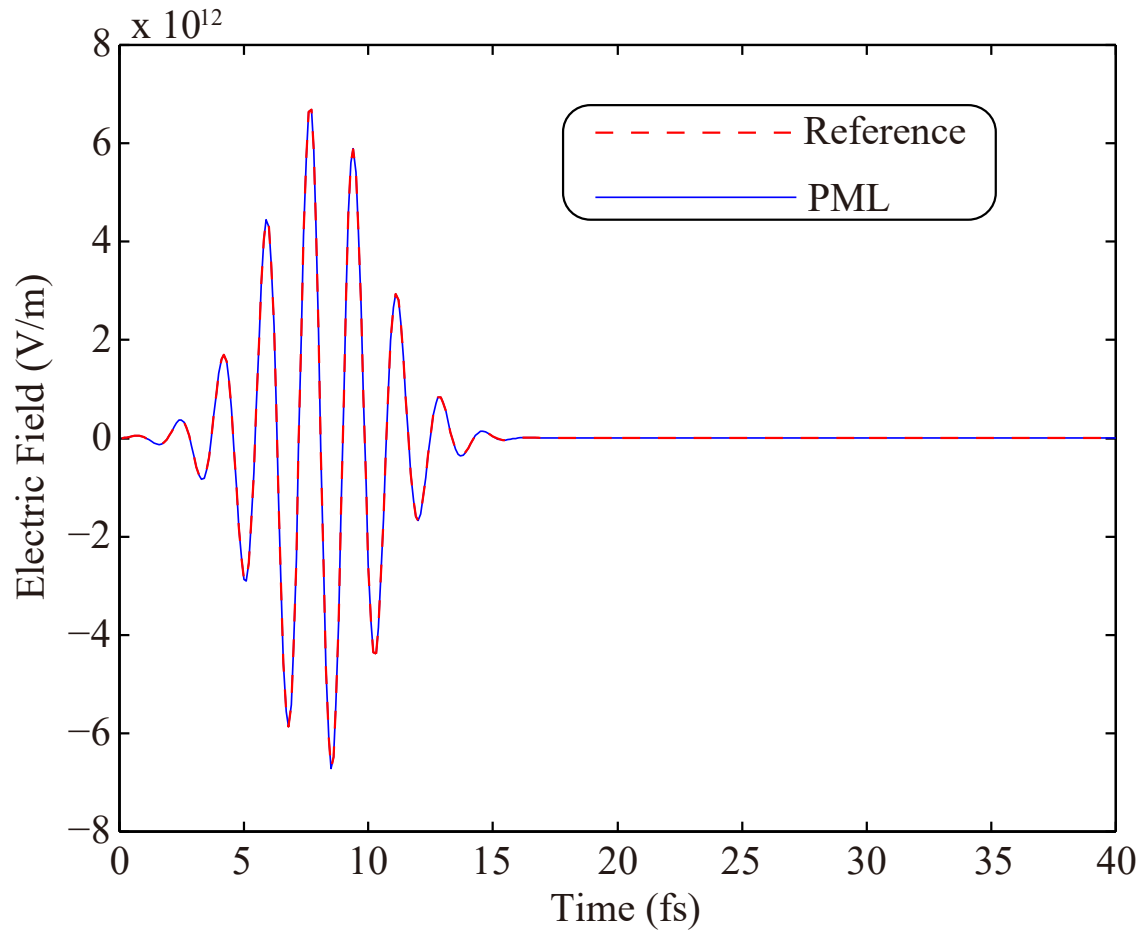

(a)

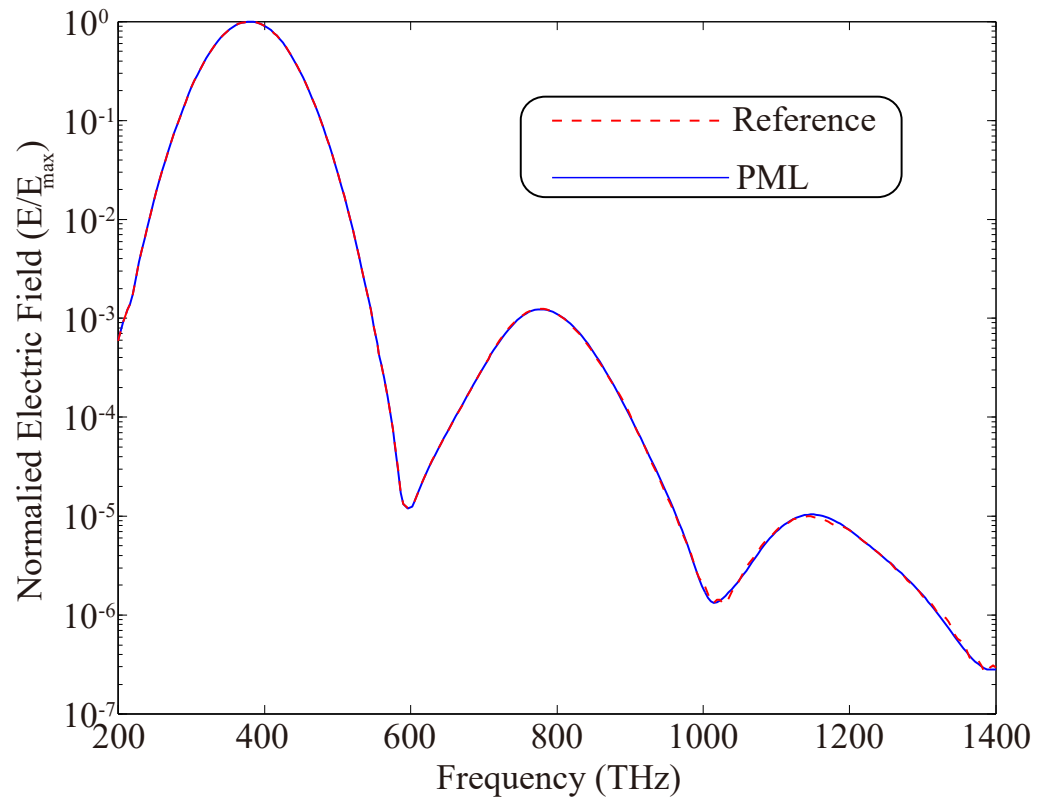

(b)

Fig. 10. (a) Electric field as a function of time. (b) A semi-log plot of the Fourier transform of the electric field as a function of frequency with the logarithmic scale used for the $y$ axis.

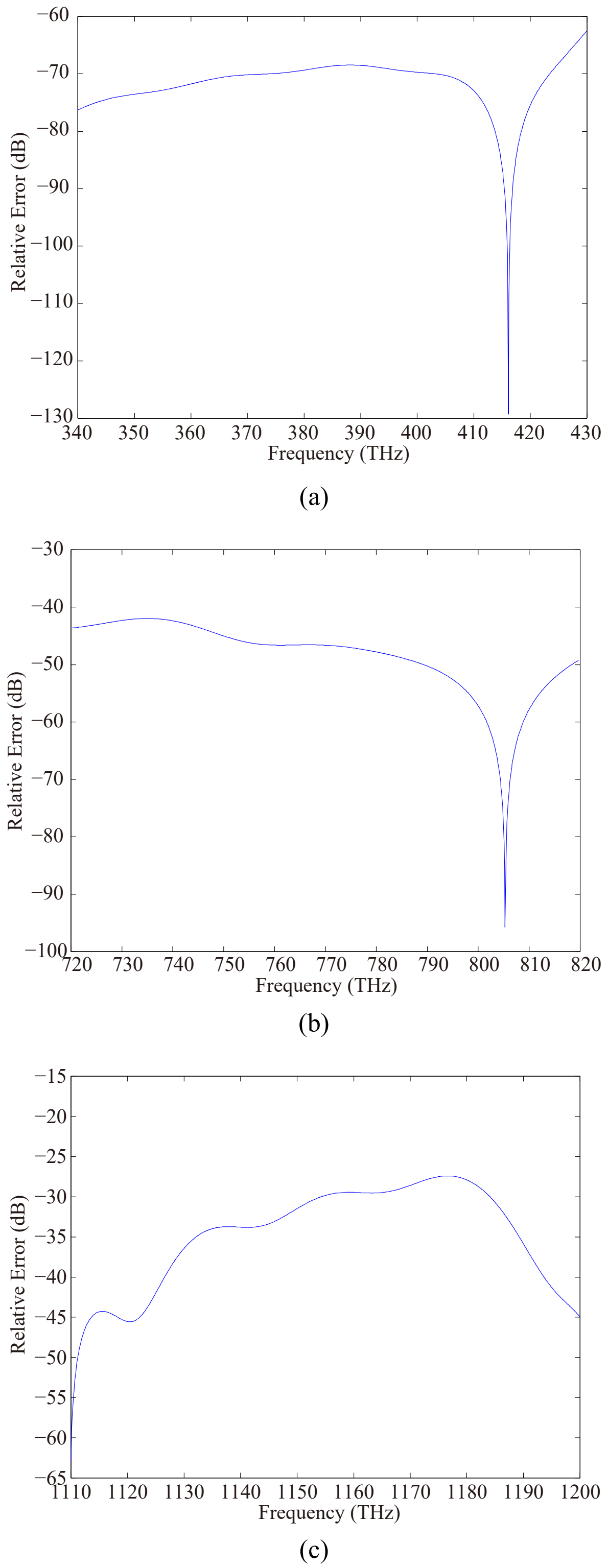

Fig. 11. Reflection errors for the (a) linear scattered waves, (b) second harmonic waves and (c) third harmonic waves.

## *D. SHG for Circularly Polarized Pulses*

In this section, we numerically verified our model by calculating the nonlinear response for metasurfaces with different rotational symmetries. The results were compared to the selection rules of SHG in the **Appendix B** and experimental results [33]. In fact, the selection rules indicate angular momentum conservation in the process of SHG, which is a powerful law for analyzing the second harmonic response of metallic nanostructures with rotational symmetries.

The metasurfaces were arranged periodically in the *xoy* plane and were illuminated by a circularly polarized beam propagating along the *z* axis (i.e., symmetry axis). The cross-sections of the triangle nanorod, L-shape, and cylinder in the *xoy* plane are shown in Fig. 12. The dimensions of the unit cells are indicated in Fig. 12 also. The thicknesses of the metasurfaces were 100 nm. The PMLs were used in the *z* direction and the periodic boundary conditions were imposed at both the *x* and *y* directions. The metasurfaces have the same lattice constant of 700 nm. The incident cosine-modulated Gaussian pulse was circularly polarized and propagated in the *z* direction as shown in Fig. 12(d). The driving frequency, temporal width, and maximum amplitude were $f_0 = 380$ THz, $\tau = 20$ fs, and $E_0 = 10^{10}$ V/m, respectively. In the FDTD simulation, we used a uniform grid with $\Delta x = \Delta y = \Delta z = 1.25\times10^{-9}$ m, and a time step $\Delta t = 1.5\times10^{-18}$ s. A computational domain of 560×560×400 grids and a total of $6.0\times10^5$ iterations were used. The total computational memory was ~24.9 GB. In the FDTD model, the spatial resolution of grids is 1.25 nm, which is typically smaller than the highest spatial resolution achievable in current fabrication technique (electron-beam lithography). However, the selection rules are not susceptible to geometric distortions; and our theoretical results agree well with the experimental ones (See Figs. 2 and 3 of the reference [33]).

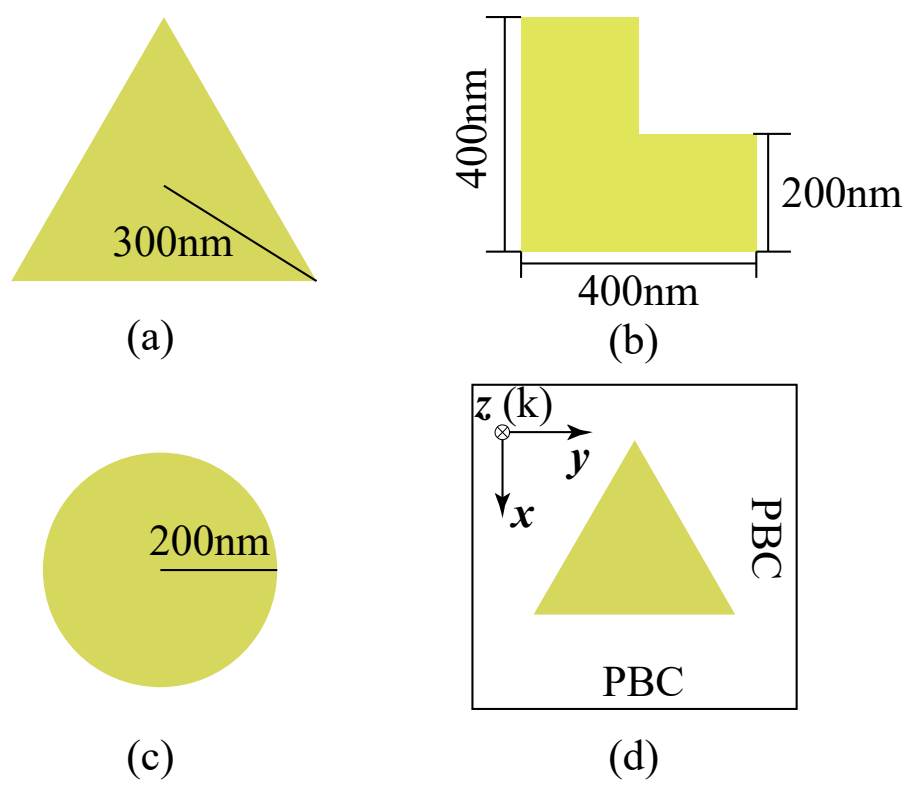

Fig. 12. Unit cells of metasurfaces with different rotational symmetries. (a) A triangle rod with a three-fold rotational symmetry. (b) An L-shape with symmetry breaking. (c) A cylinder with a central symmetry. (d) Simulation layout for modeling the SHG from the metasurfaces illuminated by circularly polarized pulses.

Excited by the circularly polarized beam, the transmitted left-circularly polarized (LCP) and right-circularly polarized (RCP) components of the second harmonics are shown in Fig. 13. The LCP and RCP components of the transmitted electric field in the FDTD method were separated by the method presented in **Appendix C**. Figures 13(a) and (b) show that for the triangle nanorod array, the polarization state of the second harmonics is almost flipped in reference to the fundamental beam. The dominant component is the RCP component under the LCP excitation, and vice versa. For the L-shaped array in Figs. 13 (c) and (d), the observed LCP and RCP components were of comparable intensity. For the cylindrical array in Figs. 13 (e) and (f), the intensities for both polarized components were quite small compared with the two other cases. From the numerical experiments, the polarization states of the second harmonic beam can be controlled by a metasurface with a predefined symmetry group. The observation is of practical importance to nonlinear devices for detection, sensing, and optical manipulation.

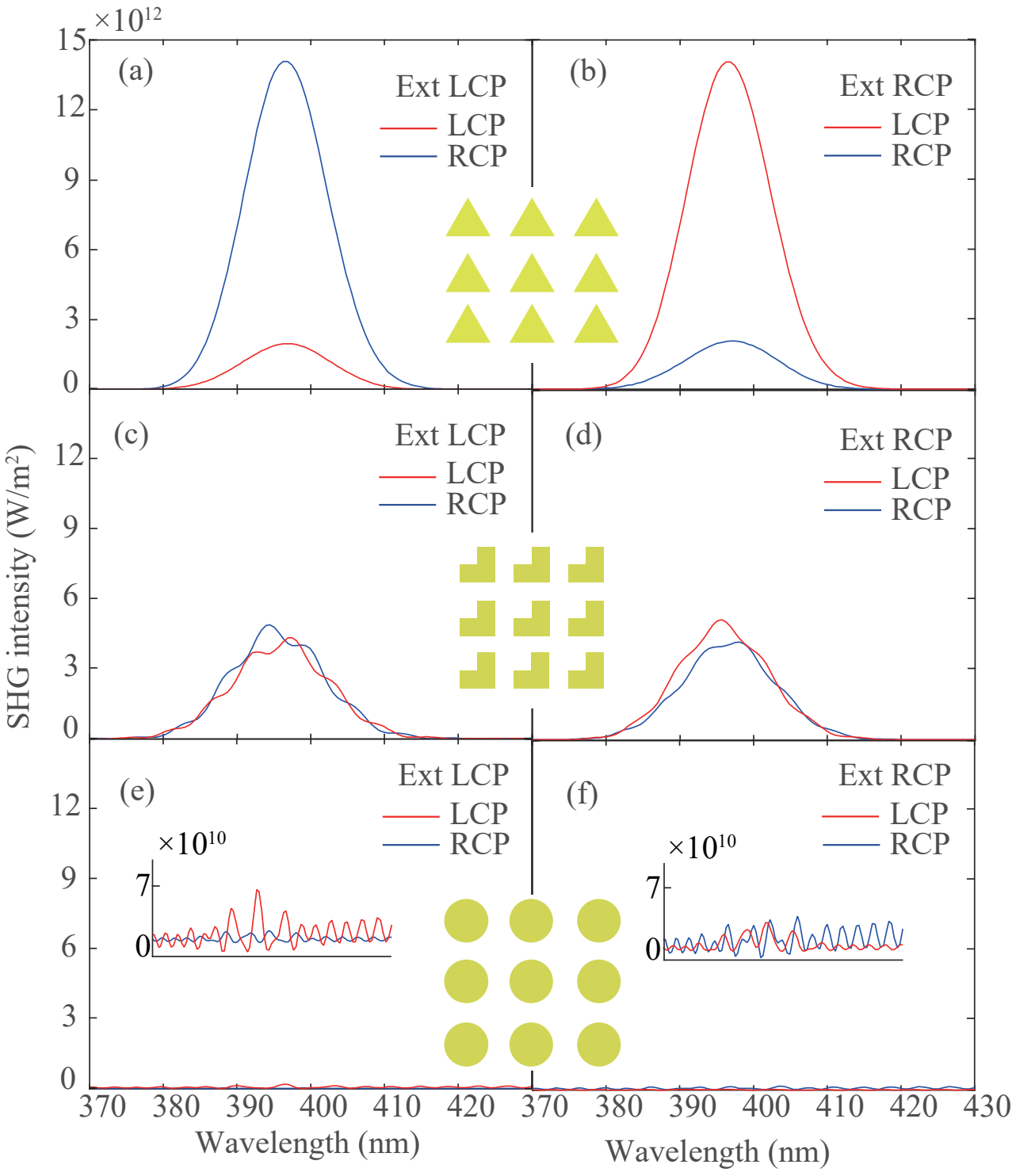

Fig. 13. Calculated second harmonic spectra for (a),(b) triangle array, (c),(d) L-shaped array and (e),(f) cylindrical array. The insets in (e) and (f) show the second harmonic intensities, magnified 50-fold. Left (right) figure shows the result for the excitation of a left-circularly (right-circularly) polarized plane wave. Red (blue) lines indicate the left-circularly (right-circularly) polarized components of the transmitted second harmonic waves.

## IV. Conclusions

In conclusion, we presented a self-consistent FDTD method for investigating both linear and nonlinear responses of metallic metamaterials. We have shown that the developed method enables to investigate nonlinear high-order harmonics from arbitrarily shaped metallic nanostructures and for different levels of pump excitation. We validated our multiphysics model by comparing it to the traditional Drude model and analytical model, and by demonstrating the exact power dependence for the second and third harmonics generated from a gold nanosphere. In particular, the polarization states of the second harmonic waves generated from metamaterials with a pre-defined rotational symmetry satisfactorily captured the corresponding selection rules. These findings are likely to be useful for the structural analysis of biological molecules [34] and spin-polarized photoemission spectroscopy [35], which require circularly polarized emitters in the vacuum-ultraviolet frequency region. In future, a more accurate solution scheme of nonlocal and quantum effects can be developed by incorporating electron exchange correlations into our approach [36].

## Appendix

### A. *Comparisons between Different Models*

To compare the hydrodynamic model to other existing schemes, the second-harmonic radiation pattern for a 100 nm radius gold sphere was calculated using the BEM. Figure 14 shows the calculated patterns. The sphere was illuminated by a plane wave at the frequency of 385 THz. The main deviations between the BEM and FDTD method lie at the side lobes of the second-harmonic radiation patterns. Regarding the centrosymmetric metals, the BEM only considers the dominated surface nonlinear sources and ignores the small bulk nonlinear sources for the SHG.

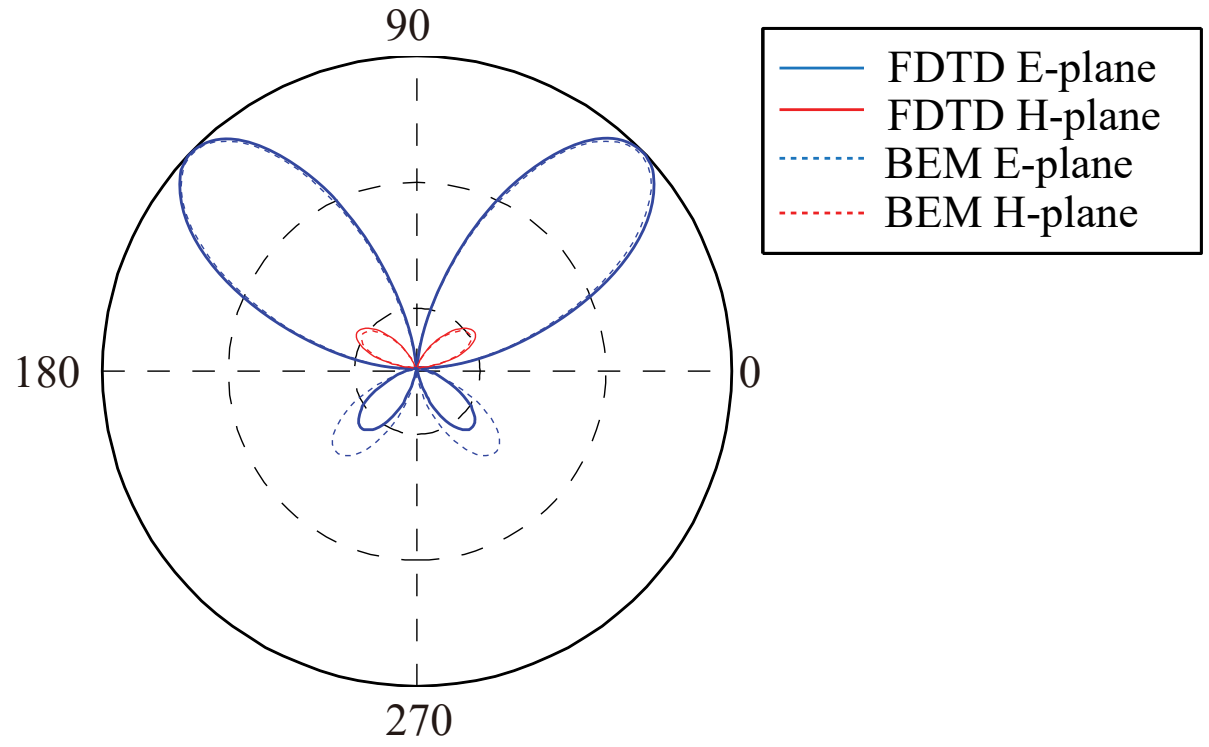

Fig. 14. Second-harmonic radiation pattern of a 100 nm radius gold sphere calculated by the boundary element method (BEM) and the proposed FDTD method. The computational configuration for the BEM can be found in our published work [17].

### B. *The SHG Selection Rule*

In this appendix, we present a powerful principle for analyzing the second harmonic response of nanostructures, and we use it to analyze the selection rules for a circular framework.

In the treatment, both second harmonic and fundamental waves propagate in the $z$ direction. The normal of a $N$-fold rotational symmetric metasurface points in the $z$ direction. In the Cartesian basis, an electric field can be written as $\mathbf{E}=E^x\boldsymbol{e}_x+E^y\boldsymbol{e}_y+E^z\boldsymbol{e}_z$. The electric field of the transverse wave in the circular frame can be given by

$$\mathbf{E}(\omega)=\tilde{E}^l(\omega)\mathbf{e}_l+\tilde{E}^r(\omega)\mathbf{e}_r+\tilde{E}^z(\omega)\mathbf{e}_z \quad \text{(B1)}$$

Here $l=+1$, $r=-1$, and $z=0$ represent the left-circular, right circular and normal components at a frequency $\omega$, respectively. The left-circularly and right-circularly polarized waves have the unit vectors $\boldsymbol{e}_l = 1/\sqrt{2}(\boldsymbol{e}_x + i\boldsymbol{e}_y)$ and $\boldsymbol{e}_r = 1/\sqrt{2}(\boldsymbol{e}_x - i\boldsymbol{e}_y)$. The electric field of the second harmonics at the frequency $2\omega$ can be obtained from the incident electric field at $\omega$

$$E^{\nu}(2\omega) = \sum_{\mu\rho=\pm 1} \Lambda^{\nu}_{\mu\rho} E^{\mu}(\omega) E^{\rho}(\omega) \qquad \text{(B2)}$$

where $\Lambda^{\nu}_{\mu\rho}$ is the nonlinear response tensor. If the coordinates are rotated by the angle $\phi$ around the $z$ axis, the rotated tensor is given by

$$\Lambda^{\nu'}_{\mu'\rho'}(\phi) = \delta^{\nu'}_{\nu}(\phi)\delta^{\mu'}_{\mu}(\phi)\delta^{\rho'}_{\rho}(\phi)e^{i(-\nu+\mu+\rho)\phi}\Lambda^{\nu}_{\mu\rho}. \qquad \text{(B3)}$$

According to Neumann's principle and Eq. (B3), in the $N$-fold symmetry $\Lambda^{\nu}_{\mu\rho}$ is invariant with the rotation angle $\phi = 2\pi / N$. Then, the following must hold

$$-\nu + \mu + \rho = nN. \qquad \text{(B4)}$$

where $n$ is an integer. This is the polarization selection rule for the second-harmonic process, for a metasurface with the $N$-fold rotational symmetry. For a fundamental incident wave propagating in the $z$ direction, the parameters $\nu$, $\mu$, and $\rho$ are either +1 or -1. If the unit cell of the metamaterial has three-fold rotational symmetry $N$=3, only the combinations $\nu$=+1, $\mu$=-1, $\rho$=-1and $\nu$=-1, $\mu$=+1, $\rho$=+1 are allowed. For $N$>3, no combination of $\nu$, $\mu$, $\rho$ is allowed and SHG is forbidden. Thus, for the fundamental wave propagating along the rotational symmetry axis, the SHG is forbidden for the centrosymmetric cylindrical structures.

In addition, in a mirror symmetric structure, we assume that the mirror plane is the $x$-$z$ plane. The covariant unit vector operator at the mirror position of the unit vector $\boldsymbol{e}_\nu$ yields

$$\boldsymbol{e}_{\nu'} = -\boldsymbol{e}_{\nu}. \qquad \text{(B5)}$$

Thus, the SHG tensor $\Lambda^{\nu}_{\mu\rho}$ in the new system of coordinate is given by

$$\Lambda^{\nu'}_{\mu'\rho'} = \delta^{\nu'}_{-\nu}\delta^{-\mu}_{\mu'}\delta^{-\rho}_{\rho'}\Lambda^{\nu}_{\mu\rho}. \qquad \text{(B6)}$$

According to Neumann's principle, the restriction on the SHG tensor can be summarized as

$$\Lambda^{\nu}_{\mu\rho} = \Lambda^{-\nu}_{-\mu-\rho}, \qquad \text{(B7)}$$

Therefore, the left-circular polarized wave is equal to right-circular polarized wave.

### C. Circular Component in FDTD

For a transmitted second-harmonic wave illuminated by a $z$-directed circular excitation, complex amplitudes of electric fields in the frequency domain, $\tilde{E}_x$ and $\tilde{E}_y$, can be divided into left-circularly and right-circularly polarized components, i.e., $\tilde{E}^l_x$, $\tilde{E}^r_x$, $\tilde{E}^l_y$ and $\tilde{E}^r_y$. We assume

$$A^l + iB^l = \tilde{E}^l_x, \; A^r + iB^r = \tilde{E}^r_x. \qquad \text{(C1)}$$

In the circularly polarized frame, we have $\tilde{E}^l_x = \tilde{E}^l_y$ and $\tilde{E}^r_x = \tilde{E}^r_y$. Considering the phase delay between the left-circularly and right-circularly polarized components, the components $\tilde{E}^l_y$ and $\tilde{E}^r_y$ can be written as

$$A^l - iB^l = \tilde{E}^l_y, \; -A^r + iB^r = \tilde{E}^r_y. \qquad \text{(C2)}$$

By solving the linear Eqns. (21-22), the amplitudes of left-circularly and right-circularly polarized waves are given by

$$I^l = \left(\frac{\mathrm{Re}(\tilde{E}x) + \mathrm{Im}(\tilde{E}y)}{2}\right) + \left(\frac{\mathrm{Im}(\tilde{E}x) - \mathrm{Re}(\tilde{E}y)}{2}\right) \qquad \text{(C3)}$$

$$I^r = \left(\frac{\mathrm{Re}(\tilde{E}x) - \mathrm{Im}(\tilde{E}y)}{2}\right) + \left(\frac{\mathrm{Im}(\tilde{E}x) + \mathrm{Re}(\tilde{E}y)}{2}\right). \qquad \text{(C4)}$$